\documentclass[aps,prl,twocolumn,superscriptaddress]{revtex4}
\usepackage{subfigure,epsfig}

\begin{document}


\title{Driving forces for Ag-induced periodic faceting of vicinal Cu(111)}

\author{A. R. Bachmann}
\email[Corresponding author. Electronic address:
]{{andreas.bachmann@physik.uni-osnabrueck.de}}
\affiliation{Fachbereich Physik, Universit\"at Osnabr\"uck,
Barbarastrasse 7, 49069-Osnabr\"uck, Germany}
\affiliation{Nijmegen Research Institute of Materials, University
of Nijmegen, Toernooiveld 1, 6525 ED Nijmegen, The Netherlands}
\author{S. Speller}
\affiliation{Nijmegen Research Institute of Materials, University of Nijmegen, Toernooiveld 1, 6525 ED Nijmegen, The Netherlands}
\author{A. Mugarza}
\affiliation{Departamento de F\'{\i}sica Aplicada I,  Universidad del Pa\'{\i}s Vasco, Plaza de O\~nate 2, E-20018 San Sebastian, Spain}
\author{J. E. Ortega}
\affiliation{Departamento de F\'{\i}sica Aplicada I,  Universidad del Pa\'{\i}s Vasco, Plaza de O\~nate 2, E-20018 San Sebastian, Spain}
\affiliation{Donostia International Physics Center and Centro Mixto CSIC/UPV, Manuel Lardizabal 4, 20018-San Sebastian, Spain}

\date{\today}

\begin{abstract}
Adsorption of submonolayer amounts of Ag on vicinal Cu(111) induces
periodic faceting. The equilibrium structure is characterized by
Ag-covered facets that alternate with clean Cu stripes. In the atomic
scale, the driving force is the matching of Ag(111)-like packed rows
with Cu(111) terraces underneath. This determines the preference for
the facet orientation and the evolution of different phases as a
function of coverage. Both Cu and Ag stripe widths can be varied
smoothly in the 3-30 nm range by tuning Ag coverage, allowing to test
theoretical predictions of elastic theories.
\end{abstract}


\maketitle

Self-organization on crystal surfaces is a promising alternative
for growing uniform nanostructures with regular sizes and spacings
\cite{Pohl1Plass1Noetzel1Brune1,Shchukin1}. In the case of
one-dimensional nanostructures one can use the regular step arrays
of vicinal surfaces as templates. Ideally atomic steps are
preferential adsorption sites, and wires and stripes can be grown
by step-flow. However, if the terrace width increases to a few
nanometers, single steps tend to meander and their spacing becomes
irregular. Yet vicinal surfaces offer another alternative at this
mesoscopic scale, i.e. using adsorbates to induce the formation of
periodic facets, which are much stiffer than individual steps. In
this case the miscut angle as well as the adatom concentration can
be tuned to obtain a variety of stripe patterns that are potential
templates for further nanostructure growth
\cite{Seehofer1Aoki1Minoda1,Himpsel1,Minoda2Horn1,Bachmann1,Foelsch1}.
The fundamental problem is controlling this complex process in
order to predict and obtain useful template structures. Therefore
it is important to understand the microscopic mechanisms that
determine the stripe pattern formation.

In the atomic scale, the driving force of the adsorbate-induced
faceting appears to be the lattice matching of the adsorbate layer
with a particular substrate orientation. The high step density of
the vicinal surface facilitates the step bunching and the
formation of facets with better lattice matching in the direction
perpendicular to the steps. Thus at a temperature where both
adatom and surface step mobilities are high enough the system
breaks up into two phases, namely well-matched adsorbate-covered
facets and adsorbate-free stepped stripes. A widely studied case
is Au on vicinal Si (v-Si). Here a variety of overlayer
reconstructions match to different Si crystal planes, leading to a
family of striped patterns made of low symmetry facets and wide
(111) or (100) terraces with different Au concentrations
\cite{Seehofer1Aoki1Minoda1,Himpsel1,Minoda2Horn1}. The final
morphology depends on the total amount of Au deposited, the
temperature and the vicinal angle of the clean surface.
Furthermore, one can tune miscut angle and Au coverage to obtain a
well-matched single phase system \cite{Himpsel1}. More recently
adsorbate-induced faceting and nanostructuration has been observed
for Ag and NaCl on v-Cu(111) \cite{Bachmann1,Foelsch1}. In these
cases the lattice matching is good between Ag(111) and Na(100)
packed layers and Cu(112) and Cu(223) planes, respectively. This
induces stripe patterns of covered and uncovered facets, as well
as changes in the facet orientation as a function of coverage.

The important question concerns the mesoscopic scale, i.e. the
nanostructure periodicity. Using the elastic theory of continuous
media for two-phase systems Marchenko has shown that the
difference in stress between the two phases produces long-range
interactions, leading to periodic domain patterns
\cite{Marchenko1}. Although this theory explains well the
spontaneous faceting of clean surfaces \cite{Rousset1}, or the
coverage-dependent periodicity of adsorbate patterns on flat
surfaces \cite{Vanderbilt1}, it is an open question whether it can
be applied for adsorbate-induced faceting.  On one hand, because
the growth of strained layers might bring in other types of
elastic relaxations, thereby suppressing the characteristic
wavelength \cite{Tersoff1}. On the other hand, the
adsorbate-induced faceting is a more complex process that involves
huge mass transport and restructuration of two different chemical
species, and hence it is more likely to be kinetically restricted.
This is the case of Au/v-Si(100) and NaCl/v-Cu(111)
\cite{Foelsch1,Meyer1}. Here we analyze the Ag-induced faceting of
v-Cu(111). We show that this self-assembled striped structure can be
tuned in the 3-30 nm range by simply controlling Ag coverage \cite{Bachman2}.
The morphology of the system is stable over a wide
temperature range. At the atomic scale we show that faceting is
driven by the tendency of Ag to form (111)-like packed layers with
minimum stress. At the mesoscopic scale we are able of proving,
for the first time in adsorbate-induced faceting, the
coverage-dependent periodicity predicted by the continuum elastic
theory at thermal equilibrium.

The substrate is v-Cu(111) with $11.8^{\circ}$ miscut about the
[$\bar{1}\bar{1}$2] direction. The surface preparation is
described elsewhere \cite{Bachmann1}. Ag deposition was done with
the substrate held at 300K. The regular stripe structure is
produced after subsequent annealing to 420K. Further annealing to
700K does not produce any appreciable change in Scanning Tunneling
Microcopy (STM) images or Low Electron Diffraction (LEED)
patterns, strongly suggesting thermal equilibrium. In order to
have a continuous variation of the coverage, Ag is deposited as a
wedge. The coverage is calculated from STM images using 3000
\AA$^2$ frames and assuming that Ag covered areas consist of
packed, 1 monolayer (ML) thick patches. Such assumption is
consistent with the structural analysis of Ag-covered areas and
also with Auger Electron Spectrocopy measurements
\cite{Bachmann1}.

In Figures 1 (a) to 1 (e) we study the morphology of the system as
a function of Ag coverage by means of STM and LEED. Fig. 2
schematically depicts the side view of the faceted structure
deduced from the analysis. The clean surface shows a regular array
of monoatomic, \{100\}-like steps along the [$\bar{1}$10]
direction, giving rise to the characteristic spot splitting in
LEED. The average terrace width measured with both
LEED and STM is $d$=10.2 \AA, i.e. the surface mostly shows
(223)-like terraces (4$\frac{2}{3}$ atomic rows per terrace,
$d$=10.5 \AA) with a few (335)-like terraces (3$\frac{2}{3}$
atomic rows, $d$=8.4 \AA). As shown in Figs. 1(b)-(d), upon Ag
adsorption the system undergoes faceting. Below 1 ML we
distinguish three different regimes, which differ on the
crystallographic orientation of the Ag-covered facet. These are
the $A$-regime up to $\sim$0.52 ML, the $B$-regime between
$\sim$0.52 ML and $\sim$0.76 ML, and the $C$-regime from
$\sim$0.76 ML to 1 ML.

Figs. 1 (b)-(c) correspond to the $A$-regime, which is
characterized by a fairly regular hill-and-valley structure of
Ag-covered facets oriented in the (112) direction and clean Cu
stepped areas. The inset of Fig. 1 (b) shows a closer view of the
surface, where we can observe the Moire pattern in the Ag-covered
facet, and monoatomic steps at clean Cu stripes with their typical
frizzy edges. In the LEED pattern we have both the spot structure
from the Ag Moire and the splitting from the step array in the Cu
stripe. As observed in Figs. 2 (b)-(c) and depicted in Fig. 2,
increasing the Ag coverage in $A$-regime results in a higher
density of Ag-covered stripes and a reduction of the step density
at clean Cu bands. The latter is due to the fact that Cu(112)
facets ($d$=6.25 \AA, 2$\frac{2}{3}$ atomic rows per terrace) have
higher step density than the average surface, thus their relative
growth requires additional steps from Cu stripes. Such effective
step removal from Cu stripes is nicely followed in the LEED
pattern in Fig. 1 (e). The Cu splitting smoothly shrinks as a
function of Ag coverage, and both spots merge into a single one
with 0.52 ML, when $A$-regime saturates \cite{note0}.

The $B$-regime from 0.52 ML to 0.76 ML is characterized by a
smooth transition in the Ag-covered facet orientation from (112)
to (335). The latter is closer to the surface normal, allowing a
higher Ag saturation. (335)-covered areas develop directly from
the center of preexisting (112) facets or arise by melting two
contiguous (112) facets. In both cases, the (335) orientation
characterizes the center of the Ag-covered stripes, whereas the
boundaries preserve the (112) orientation. For 0.72 ML in Fig. 1
(d) the $B$-regime is almost saturated, i.e. we have flat Cu(111)
terraces and mostly (335)-oriented Ag stripes, also displaying a
Moire structure. During the $C$-regime, from 0.76 ML to 1ML, (223)
facets develop. With 1 ML we observe a close-packed Ag layer
wetting a random distribution of (335)- and (223)-like terraces
with the same step density than the starting surface.

In the atomic scale, the driving force for the Ag-induced faceting
of the stepped Cu surface is the matching between Ag(111)-like
packed rows along the parallel direction of the steps and the Cu
facet underneath. Lattice-matching has been also proposed as the
microscopic mechanism for the NaCl-induced faceting of Cu(112)
\cite{Foelsch1}. Our system generally displays a clear tendency to
avoid adsorption on wide (111) terraces, which leads to a large
mismatch \cite{Meunier1}. On the other hand, as deduced from Figs.
1 and 2, there is a preference for (112)-oriented facets at low
coverages, in spite that this requires a larger mass transport.
This can be naively explained from the width of the Cu terrace
required for a good matching in the perpendicular direction. Five
Ag packed rows (2.498 \AA\ wide each) fit to two 6.25 \AA\ wide
terraces in Cu(112) with only 0.12\% mismatch in the direction
perpendicular to the steps, whereas Cu(335) requires two 8.37 \AA\
wide terraces to accommodate seven Ag rows with 4.47\% mismatch.
This preference for (112) facets is also supported by the
microscopic analysis of (112) and (335) stripes shown in Fig. 3.
The atomic models in the right panels of Fig. 3 reproduce the
Moire patterns of the left panels in the simplest way. The open
circles represent Ag-close packed layers, the small dots the
Cu(112) and Cu(335) unit cell underneath, and the shades the
"on-top" positions. Thus, in this simple approach the step
corrugation is being disregarded. We use the smallest compressions
and azimuthal rotations of the Ag(111) adlayer to obtain the
Moires that fit to the STM observations. For both (112)- and
(335)-oriented layers the Ag layer rotation is the same
5.2$^{\circ}$, allowing the smooth transition from (112) to (335)
facets observed beyond 0.52 ML. For (112) facets the Ag adlayer is
compressed by 0.5\% and 7.8\% in the direction perpendicular and
parallel to the rows, respectively. For (335) facets the
perpendicular and parallel compressions are 6.2\% and 7.8\%,
respectively. Thus, the (335)-oriented layer accumulates much more
stress, explaining the preference to form (112)-oriented stripes.

The long range periodicity of the Ag-induced faceting of Fig. 1 is
nicely explained within Marchenko's elastic theory, as shown in
Fig. 4. Data points represent the nanostructure period ($L$) and
the Ag stripe width ($w_{Ag}$) measured from the STM images as a
function of coverage ($\theta$) from 0.1 ML to 0.9 ML. The
periodicity is poorly defined above 0.52 ML, and in this case we
consider the average value $\langle L \rangle$, defined as
$\langle L \rangle=\langle w_{Ag} \rangle + \langle w_{Cu}
\rangle$, where $\langle w_{Ag} \rangle$ and $\langle w_{Cu}
\rangle$ respectively are the average widths of Ag-covered facets
and Cu stripes. The lines in Fig. 4 are fit to the data using
Marchenko's model, where the superlattice periodicity $L$ and the
stripe width $w_{Ag}$ are given by \cite{Marchenko1}:

\begin{equation}
L(\theta) = \frac{\kappa}{\sin(\pi\theta)} \;\;\; ; \;\;\;
w_{Ag}(\theta) = \frac{\kappa \theta}{\sin(\pi\theta)}
\end{equation}

$\kappa$ is the only adjustable parameter. The model is strictly
defined for a two-phase system, and hence the fit is restricted to
the $A$-regime. The result is excellent and we obtain $\kappa_A$ = 91$\pm$6 \AA.
Data points at regime transitions display a clear  discontinuity.
Microscopically, this appears related to the way some new facets
develop, i.e. by melting two contiguous facets in the previous
regime. But, as indicated in Fig. 4,
by changing in Eq. 1 to $\kappa_{B}$= 110$\pm$8  \AA\ and $\kappa_{C}$= 141
$\pm$8 \AA\ both $\langle w_{Ag}(\theta) \rangle$ and $\langle L
(\theta)\rangle$ in $B$- and $C$-regimes are still nicely reproduced.

In Marchenko's approach $\kappa$ is related to elastic constants
via $\kappa= 2 \pi \; a \times {\rm exp}(1+C_1/C_2)$, where $a$ is
a microscopic cutoff and $C_1$ is the free energy of the phase
boundary per unit length \cite{Marchenko1}. If we assume that work
function variations between Cu and Ag phases are relatively small
\cite{Vanderbilt1}, $C_2$ has only elastic nature and reflects the
stress difference between phases \cite{Marchenko1}:

\begin{equation}
C_2 = \frac{1-\nu^2}{\pi E} \; |\vec{F}|^2
\end{equation}

where $E$ and $\nu$ respectively are the shear modulus and the
Poisson ratio for Cu, and
$\vec{F}=(\vec{\sigma}_{Ag}-\vec{\sigma}_{Cu})$ the force exerted
at facet edges due to the difference in surface stress between
Ag-covered facets and Cu stripes, as defined in the inset of Fig.
4. The general agreement with Marchenko's theory shown in Fig. 4
indicates that $\kappa$ in Eq. 1 does not have a significant
coverage-dependence. The $\theta$-dependence is expected in the
elastic constants due to the changing shape of Cu stripes and
Ag-covered facets. Even with coverage-dependent elastic constants
Eq. 1 always holds if there is a (relatively) high stress energy
in any of the Ag-covered facets, such that $C_1$/$C_2\sim 0$. The
actual value of $C_1/C_2$ can be obtained from the fit in Fig. 4
assuming a cutoff length $a$. The cutoff length is defined as the
minimum size of the (growing) domain \cite{Vanderbilt1,Men1},
which in this context is the terrace width. Using $\kappa_A$ =
91$\pm$2 \AA\ and $a$=6.25 \AA, we indeed obtain $C_1/C_2$= -0.15
$\pm$ 0.05. The absolute value of $C_1/C_2$ is thus much smaller
than any other found in the literature, and can be assumed to be
zero \cite{note3}. The condition $C_1$/$C_2\sim 0$ is expected to
hold for $B$ and $C$ regimes as well. The boundary energy must be
very similar, because the atomic structure at facet edges is the
same in all regimes, whereas the stress difference is expected to
increase, since the preference for (112) orientation indicates
that the lowest surface energy occurs for this facet. Thus only a
change in $a$ would explain the changes in $\kappa$ in Fig. 4. The
discontinuities in Fig. 4 are in fact consistent with the
increasing terrace size (cutoff $a$) of the growing facet. The
fitting parameter variation from $\kappa_A$ = 91 \AA\ to
$\kappa_{B}$= 110 \AA\ (21\%) and to $\kappa_{C}$= 141 \AA\ (28\%)
closely correlates with the relative increase in the terrace size
from $d_{112}$=6.25 \AA\ to $d_{335}$=8.4 \AA\ (34\%) and to
$d_{223}$=10.5 \AA\ (25\%). Since $\kappa$ discontinuities were
microscopically related to melting of contiguous facets at regime
transitions, this makes the real nature of the cutoff length an
interesting issue that encourages a detailed microscopic model.

Note that the $C_1$/$C_2\sim 0$ case leads to the shortest possible
superlattice period in Ag/v-Cu(111). This contrasts with the long
superlattice periods observed at high temperatures in
NaCl/v-Cu(111) \cite{Foelsch1}. Long periods are only possible if
$C_1/C_2\gg 1$, which in the latter case could be due to a more
efficient interface relaxation, favored by a large
adsorbate-surface charge transfer that weakens Cu-Cu bonds.

A.R.B. and S.S. are supported by the Deutsche
Forschungsgemeinschaft (DFG) and the Stichting voor Fundamenteel
Onderzoek der Materie (FOM). A.M. and J.E.O. are supported by the
Universidad del Pa\'{\i}s Vasco
(1/UPV/EHU/00057.240-EA-8078/2000)and the Max Planck Research
Award Program. Fruitful discussions with A. Rubio are
acknowledged.

\bibliographystyle{aipbib}

\begin{thebibliography}{1}
\bibitem{note0} $A$-regime ideally saturates at 0.61 ML,
when Cu stripes become flat (111) terraces \cite{Bachmann1}.
However some steps appear pinned by defects within Cu stripes,
such that the effective saturation coverage is lower.
\bibitem{note3}. With such a small $C_1/C_2$ ratio, the negative sign, which is not physical, must be attributed to the elastic model limitations.
\end{thebibliography}

\begin{figure}
\epsfig{file=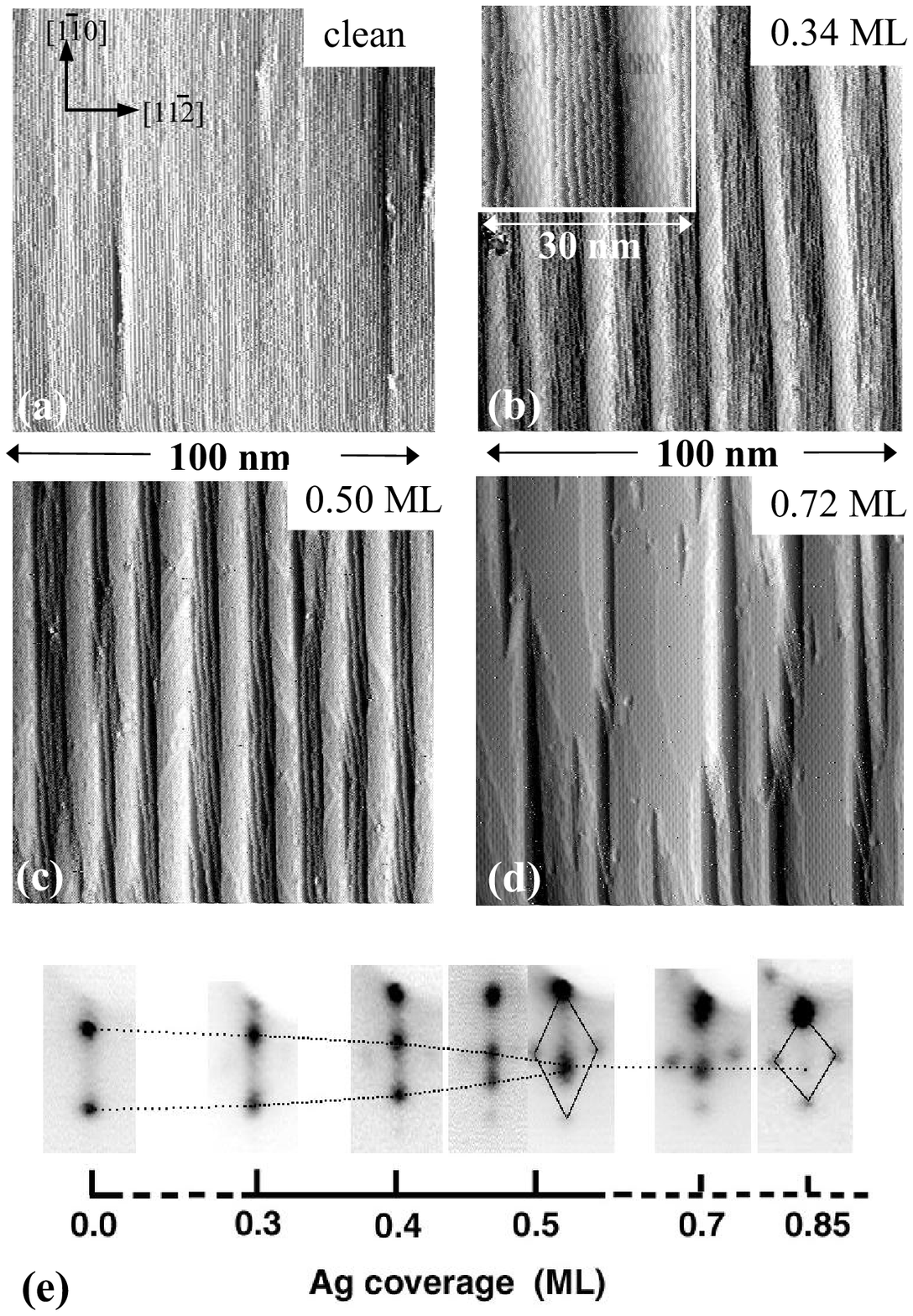,width=1.0\linewidth}
\caption[]{(a)-(d) STM images and (e) (0,0) LEED spot measured at
different coverages for Ag-induced faceting in vicinal Cu(111). Up
to 0.6 ML, the system displays periodic, two-phase separation of
Ag-covered (112) facets and clean Cu stepped stripes. Both give
distinct LEED structures in (e), i.e. spot splitting for Cu
stripes and Moiree pattern spots for Ag facets. As the Ag coverage
increases from (a) to (c), steps from Cu stripes incorporate in Ag
facets, such that Cu terraces become wider. This is proved by the
splitting reduction in (e). (335) facets like those in (d) develop
beyond 0.6 ML.} \label{fig1}
\end{figure}

\begin{figure}
\epsfig{file=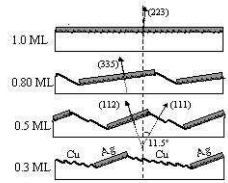,width=0.8\linewidth,clip=}
\caption[]{Schematic evolution of the periodic faceting induced by
Ag on vicinal Cu(111) with $11.8^{\circ}$ miscut, as deduced from
the STM-LEED analysis in Fig. 1.} \label{fig2}
\end{figure}

\begin{figure}
\epsfig{file=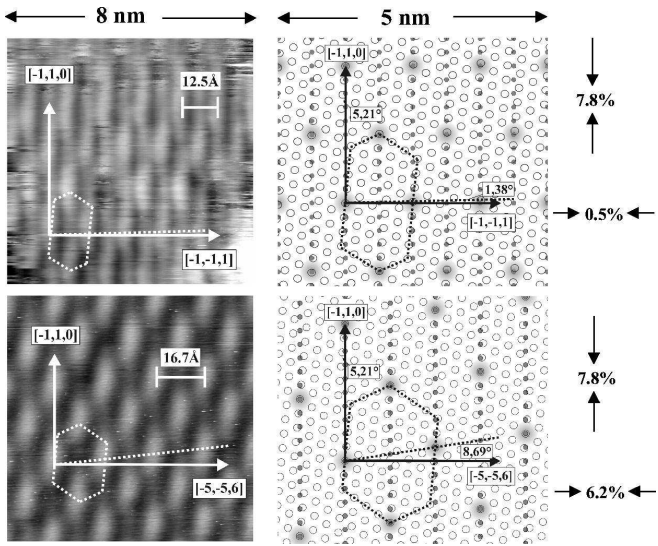,width=1.0\linewidth,clip=}
\caption[]{Left, detailed STM view of the Moiree patterns for
(112) and (335) Ag-covered facets. Right, two-dimensional atomic
models that reproduce the Moirees in the left. The small filled
circles indicate the step edge Cu atom positions underneath. Open
dots represent Ag atoms and shaded circles indicate "on-top"
positions. The fit requires the indicated compressions and
azimuthal rotations of the Ag packed layer.} \label{fig3}
\end{figure}

\begin{figure}
\epsfig{file=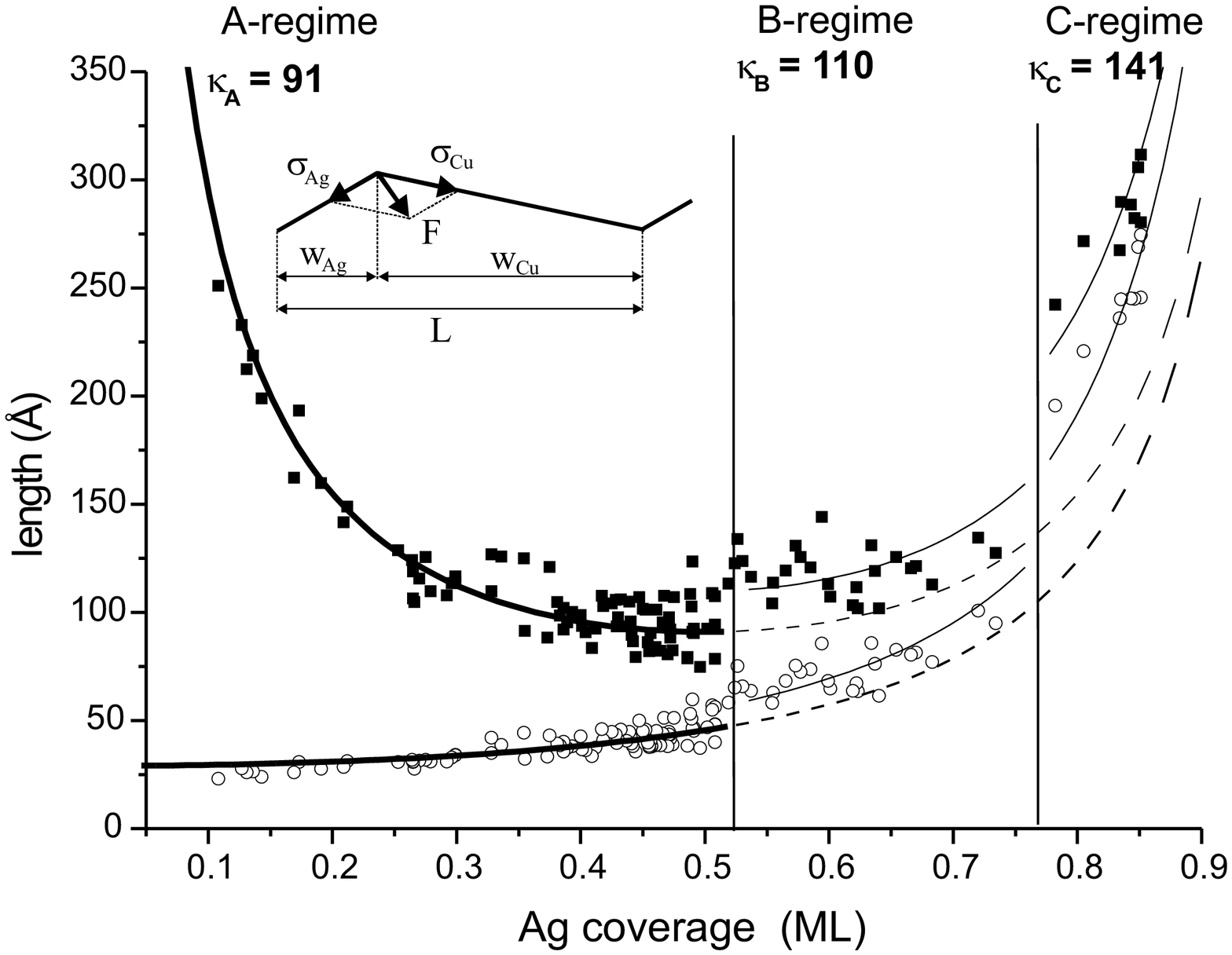,width=1.0\linewidth,clip=}
\caption[]{Nanostructure periodicity (filled squares) and Ag-covered stripe
width (open circles) as a function of Ag coverage. Vertical lines
distinguish the $A$, $B$ and $C$ regimes, which differ on the type of
the growing Ag-covered facet. The thick solid lines are fit to the data
in the two-phase $A$-regime using the continuum elastic theory of
Marchenko's \cite{Marchenko1}. A continuation in the $B-C$ regimes is represented
by the thin dotted line. The thin solid lines do fit separately data
points in the $B$ and $C$ regime. The inset shows the surface stress vectors
in a faceted structure (see the text)} \label{fig4}
\end{figure}

\end{document}